\documentclass{PHYEAUTH}
\usepackage{graphicx}
\usepackage{amsmath}
\usepackage{amssymb}

\usepackage{latexsym}
\usepackage{bm}
\newcommand{\ket}[1]{|n(#1)\rangle}
\newcommand{\braket}[2]{\langle n(#1)|n(#2)\rangle}

\newcommand{\hatmu}{\hat\mu}

\newcommand{\rmB}{{\rm B}}

\newcommand{\U}{\mathop{\rm {}U}}
\newcommand{\cal}{\rm }

\begin{document}

\begin{frontmatter}

\title{
Topological Description of (Spin) 
Hall Conductances on Brillouin Zone Lattices:
Quantum Phase Transitions and Topological Changes
}

\author[address1]{Y. Hatsugai
},
\author[address2]{T. Fukui}
and
\author[address3]{H. Suzuki}

\address[address1]{
 Department of Applied Physics, University of Tokyo, 7-3-1 Hongo, Bunkyo-ku, Tokyo 113-8656, Japan
}

\address[address2]{
Department of Mathematical Sciences, Ibaraki University, Mito 310-8512, Japan
}

\address[address3]{
Institute of Applied Beam Science, Ibaraki University, Mito 310-8512, Japan
}
\begin{abstract}
It is widely accepted that topological quantities are useful 
to describe quantum liquids in low dimensions.
The (spin) Hall conductances are typical examples. 
They are expressed by the Chern numbers, which are topological
invariants given by the Berry connections of the ground states. 
We present a topological  description for the (spin) Hall conductances 
on a discretized  Brillouin Zone.
At the same time, it is  quite efficient in practical numerical 
calculations for concrete models.
We demonstrate its validity in a model with quantum phase transitions.
Topological changes supplemented with the transition
is also described in the present lattice formulation.
\end{abstract}

\end{frontmatter}

\section{Introduction}
\vskip -0.3cm
Topological quantities are fundamental to describe low dimensional
quantum liquids where standard symmetry breaking do  not have 
a primal  importance \cite{Wen89,Hat045}.  
Typical  examples are quantum Hall liquids \cite{TKNN,Ber84,Sim83} and
anisotropic superconductors with time-reversal symmetry breaking \cite{smf,mh}.
Recently spin Hall conductance for semiconductors 
are also  attracting much  current interest \cite{smf,mh,spin1,spin2,Hal04}. 
The (spin) Hall conductance has a characteristic geometrical meaning \cite{TKNN,Sim83}.
In some physical units, they are given by the first Chern number
of the Berry connection \cite{Ber84}.

In practical numerical calculations,
we diagonalize Hamiltonians on a set of discrete points on the Brillouin
zone (BZ). It is thus crucial to develop an efficient method of
revealing the topological property of infinite systems with continuum 
BZ from corresponding finite systems with discrete BZ.
We propose an efficient method for the calculation of  the Chern numbers
on a discretized BZ based on  a geometrical
formulation of topological charges in lattice gauge
theory \cite{Luscher,Phillips,Fujiwara:2000wn}.
The Chern numbers thus obtained are {\it manifestly gauge-invariant\/} 
and {\it integer-valued\/} even for a discretized BZ.
One can compute the Chern numbers using wave functions in
{\it any gauge\/} or {\it without specifying gauge fixing-conditions}.
Details of the formulation and the basic results were published
elsewhere~\cite{fukui05}.

\section{Topological Description of (Spin) Hall conductances}
\vskip -0.3cm
\underline{Chern Numbers as (Spin) Hall conductances}:
Let us consider  Chern numbers in the quantum Hall effect as a typical
example. The spin Hall conductances is treated similarly.
We take the BZ by
$0\le k_\mu<2\pi/q_\mu$ ($\mu=1$, 2 with integers~$q_\mu$). Since the
Hamiltonian ~$H(k)$ is periodic in both $k_1$ and $k_2$ directions,
the  BZ is regarded as a two-dimensional torus~$T^2$. When the Fermi energy
lies in a gap, the Hall conductance is given by
$\sigma_{xy}=-(e^2/h)\sum_nc_n$, where $c_n$ denotes the Chern number of the
$n$th Bloch band, and the sum over $n$ is restricted to the bands below the
Fermi energy \cite{TKNN}. The Chern number assigned to the $n$th
band is defined  by 
$c_n = \frac{1}{2\pi i} \int_{T^2} F $ with $F=dA$,
where the Berry connection $A=A_\mu dk_\mu$  with 
$A_\mu=\langle n|\partial/\partial k_\mu |n\rangle$
is defined by  a normalized wave function of the $n$th Bloch band $|n\rangle$
satisfying  $H(k)\ket{k}=E_n(k)\ket{k}$ and 
$\langle n | n \rangle =1$ \cite{TKNN,Ber84,Sim83}.
The Chern number can be nonzero 
only when the gauge potential cannot be defined as a
global function over~$T^2$. In this
case, one covers $T^2$ by several coordinate patches and then, within each
patch, one can take a local gauge (a phase convention for the wave
functions) such that the gauge potential is a well defined 
function. In an overlap between two patches, gauge potentials defined on each
patch are related by a $\U(1)$ gauge transformation:
$| n \rangle \to | n \rangle \omega $ and 
$A \to  A + \omega ^\dagger  d\omega $ 
$(|\omega | =  1) $.
In the continuum theory, one needs to fix this gauge freedom to
perform any explicit calculations. 
To fix the gauge, one first selects an arbitrary state $|\phi\rangle$ 
which is globally well defined over the whole BZ \cite{Hat045}.
Then the gauge can be specified by
$ |n^\phi\rangle =  P_n|\phi\rangle/N^\phi $
(if $N^\phi\neq 0$), where $P_n=|n\rangle\langle n|$ and 
$N^\phi=|\langle\phi|n\rangle|$ are  gauge independent.
Generically, this gauge is only allowed locally
because the overlap $N^\phi$ may vanishes.
It inevitably occurs to have a non vanishing Chern number.
Then one needs to use several different state $|\phi \rangle $
 (gauges) in several regions of patches 
to cover the whole BZ.

\underline{Topological Description on a discretized BZ}:
We now switch to the finite systems with discrete BZ.
A naive replacement of the differential 
operator to the difference operator
breaks the gauge invariance and topological characters of the Chern numbers.
Here we propose an explicitly gauge invariant and topological definition 
of the Chern number on a lattice.  Let us denote lattice points~$k_\ell$
($\ell=1$, \dots, $N_1N_2$) on the discrete BZ as
$k_\ell=(k_{j_1},k_{j_2})$, $k_{j_\mu}=\frac{2\pi j_\mu}{q_\mu N_\mu}$,
$(j_\mu=0,\ldots,N_\mu-1)$.
We assume that the state $\ket{k_\ell}$ is periodic on the lattice,
$\ket{k_\ell+N_\mu\hatmu} =\ket{k_\ell}$,
where $\hatmu$ is a vector in the
direction~$\mu$ with the magnitude~$2\pi/(q_\mu N_\mu)$.
We first define a $\U(1)$ link variable from the wave functions
\begin{alignat*}{1} 
U_\mu(k_\ell) &
 \equiv {\cal N}^{-1}_\mu(k_\ell)\braket{k_\ell}{k_\ell+\hatmu},
\end{alignat*} 
where ${\cal N}_\mu(k_\ell)\equiv|\braket{k_\ell}{k_\ell+\hatmu}|$. 
It is well defined as long as ${\cal N}_\mu(k_\ell)\neq0$.
(It is always assumed  by an infinitesimal shift of the lattice.)
Next we define a lattice field strength by
\begin{alignat*}{1} 
\tilde F_{12}(k_\ell)\equiv & 
   \ln U_1(k_\ell)U_2(k_\ell+\hat1)U_1(k_\ell+\hat2)^{-1}U_2(k_\ell)^{-1},
\end{alignat*} 
($-\pi<\frac{1}{i}{\tilde F}_{12}(k_\ell)\leq\pi$).
The field strength is defined within the principal branch of the logarithm.
Finally, we define the Chern number on the lattice as 
\begin{alignat*}{1} 
   \tilde c_n &
 \equiv\frac{1}{2\pi i}
\sum\nolimits_\ell\tilde F_{12}(k_\ell).
\end{alignat*} 
We stress here  that $\tilde c_n$ is manifestly {\it gauge-invariant\/}
and 
 {\it strictly an integer\/} for
arbitrary lattice spacings.
To demonstrate it, let us introduce a gauge potential
\begin{alignat*}{1} 
   \tilde A_\mu(k_\ell) =& \ln U_\mu(k_\ell),
\end{alignat*}  
($-\pi<\frac{1}{i}\tilde A_\mu(k_\ell)\leq\pi)$
which is periodic on the lattice:
$\tilde A_\mu(k_\ell+N_\mu\hatmu)=\tilde A_\mu(k_\ell)$. 
By definition, one finds
\begin{alignat*}{1} 
   \tilde F_{12}(k_\ell)
=& \Delta_1\tilde A_2(k_\ell)-\Delta_2\tilde A _1(k_\ell)
   +2\pi in_{12}(k_\ell),
\end{alignat*} 
where $\Delta_\mu$ is the forward difference operator on the lattice,
$\Delta_\mu f(k_\ell)=f(k_\ell+\hatmu)-f(k_\ell)$, and $n_{12}(k_\ell)$ is an
{\it integer-valued\/} field, which is chosen such that
$(1/i){\tilde F}_{12}(k_\ell)$ takes a value within the principal branch.
Now we have
\begin{alignat*}{1} 
  \tilde c_n =&
 \sum\nolimits_{\ell} n_{12}(k_\ell).
\end{alignat*} 
It shows that the lattice Chern number~$\tilde c_n$ is an integer.
To avoid ambiguities we assume that there are no exceptional configurations
in the system under consideration,
$|\tilde F_{12}(k_\ell)|\ne \pi$ for all $k_\ell$ \cite{Phillips}.
This condition will be referred to as {\it admissibility}.
Under this condition, the Chern number is uniquely determined. 
Since our lattice formulation recovers the continuum one in the limit~$N_\mu\to\infty$,
we expect the lattice field strength~$\tilde F_{12}$ will
be small enough for a sufficiently large~$N_\mu$ and the lattice Chern number
will approach the one in the continuum $\tilde c_n\to c_n$ in the
this limit. Since both $\tilde c_n$ and~$c_n$ are integers, we
have $\tilde c_n=c_n$ for meshes of appropriate sizes.
As far as the system is regular,
our lattice formulation could reproduce correct Chern numbers of the continuum theory
even for a coarsely discretized BZ, $N_1N_2\sim O(|\tilde c_n|)$ \cite{fukui05}.

Also our method can be extended \cite{fukui05} 
to the case of the non-Abelian Berry connection
$A=\psi^\dagger d\psi$, which is a matrix-valued one-form
associated with a multi-dimensional multiplet $\psi$ \cite{Hat045,wz}. 

\underline{Uniqueness of the description}:
It turns out that the description presented so far proves to be 
{\it unique} for discretized BZ.
Namely, under the admissibility, the space of $\U(1)$ link variables is
divided into disconnected sectors and the topological number~$\tilde c_n$ is
uniquely assigned to each sector. 
The Chern number~$\tilde c_n$ is, moreover,
a {\it unique\/} gauge-invariant topological integer which can be assigned to
admissible $\U(1)$~link variables. 
The proof of this statement has been given by L\"{u}scher \cite{Luscher}.

A key ingredient of his proof is that
any U(1) link variable can be decomposed  uniquely into
$U_\mu(k_\ell)=e^{iA_\mu^{[\tilde c_n]}(k_\ell)+iA_\mu^{\rm T}(k_\ell)}
\Lambda(k_\ell)U^{[w]}_\mu(k_\ell)\Lambda(k_\ell+\hat\mu)$,
where $A_\mu^{[\tilde c_n]}(k_\ell)$ is a gauge field giving rise to the Chern number; 
$\Delta_1A_2^{[\tilde c_n]}(k_\ell)-\Delta_2A_1^{[\tilde c_n]}(k_\ell)
=2\pi i\tilde c_n/(N_1N_2)$,  $A_\mu^{\rm T}(k_\ell)$ 
is a periodic transverse field, 
$\Lambda(k_\ell)$ is a gauge transformation, and 
$U^{[w]}_\mu(k_\ell)$ is a gauge field giving nontrivial Wilson lines
but giving zero field strength. 
Once $A_\mu^{[\tilde c_n]}(k_\ell)$ is found, 
$A_\mu^{\rm T}(k_\ell)$ 
can be determined uniquely by  the relation
$\tilde F_{12}(k_\ell)=\Delta_1A_2^{\rm T}(k_\ell)-\Delta_2A_1^{\rm T}(k_\ell)
+2\pi i \tilde c_n/(N_1N_2)$.
Readers who are interested in the proof should refer  L\"uscher's paper.

The admissibility condition is important 
for the present lattice formulation and its breakdown is
closely related to a singular behavior of the Berry connection,
that is, it is supplemented with a quantum phase transition.
In the present context of the Berry connection, 
a distribution of the gauge invariant field $\tilde F_{12}$
is completely governed by the $k$~dependence of the Hamiltonian. 
Each of the topological ordered states with a nontrivial Chern
number corresponds to  nontrivial topological sector specified by the
admissibility and also characterized by the lattice Chern number~$\tilde c_n$.
In the continuum, on the other hand, 
the topological stability of the Chern number is assured
by the gap-opening condition \cite{Hat045}.
The topological quantum phase
transitions are thus characterized by the gap closing. Namely, nontrivial
topological sectors of the continuum, each of which is a topological ordered
state, are separated by the gaps.

In passing, we would like to mention the Kubo formula which is often 
used in numerical calculations:
$\bar{F}_{12}(k)=2i\sum_{m(\ne n)} 
\frac{{\rm Im}\langle n(k)|\partial_1 H(k)|m(k)\rangle
\langle m(k)|\partial_2 H(k)|n(k)\rangle}
{[E_n(k)-E_m(k)]^2}$.
This formula is equivalent the field strength $F_{12}(k)$ in the 
continuum BZ, but {\it not} to $\tilde F_{12}(k_\ell)$ in the discretized BZ.
In the latter case, $\sum_\ell\bar F_{12}(k_\ell)$ is no longer topological, 
although the gauge-invariance remains. 
Therefore, our topological and gauge-invariant formulation 
for discrete BZ should have strong
advantages, especially of describing regions where
topological  transitions occur.

\vspace{-3mm}
\section{Quantum  Phase Transition and Topological Change}
\vskip -0.3cm
To demonstrate the present scheme, we take the
Hamiltonian for spinless fermions in an external magnetic field:
$H=-\sum_{\langle i,j\rangle}t_{ij}c_i^\dagger e^{i\theta_{i,j}}c_j$, where
the flux per plaquette on the coordinate
lattice~$\phi=\sum_\Box\theta_{i,j}/(2\pi)$ is~$p/q$.
We consider a model with quantum phase transition, that is,
with next nearest neighbor (NNN) hopping $t'$ (nearest neighbor hopping is
$t$).
%
The Hamiltonian in the $k$-space is given by
$H_{ij}(k)=-2t\delta_{ij}\cos(k_y-2\pi\phi j)
-B_i\delta_{i+1,j}-B_j^*\delta_{i,j+1}
-B_q^*\delta_{i+q-1,j}e^{-iqk_x}
-B_q\delta_{i,j+q-1}e^{iqk_x}$,
where
$B_j=t+2t'\cos\big(k_y-2\pi \phi(j+1/2)\big)$
($i$, $j=1$, \dots, $q$) with $q_1=q$ and $q_2=1$ \cite{Hat90}.
Bellow, we will
present some results of applying our method to the middle 
subband of the $\phi=1/3$
(that is, $q=3$) system.
For simplicity, we set $N_1=N_{\rm B}$ and $N_2=qN_{\rm B}~(=3N_{\rm B})$.
\begin{figure}[htb]
\begin{center}
\includegraphics[width=0.89\linewidth]{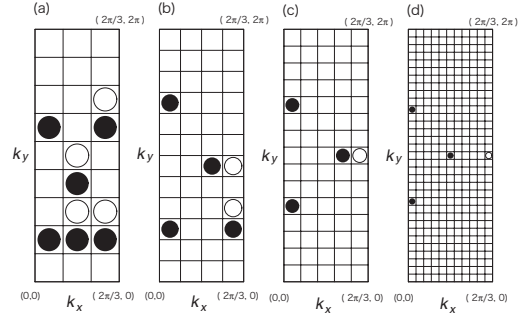}
\end{center}
\caption{Configuration of integer field~$n_{12}(k_\ell)$ of the
NN model $({t'}=0)$ in a
gauge specified by state~$|\phi_g\rangle$ over discretized
Brillouin zones.
 $N_\rmB=3(a),
4(b),
5(c)$ and
$11(d)$.
Black (white) circles denote $n_{12}=-1$ (1).}
\label{f:NFie}
\end{figure}
\vskip -0.2cm
We show the integer field~$n_{12}(k_\ell)$ 
in Figs.~\ref{f:NFie}(a)-(d), where we use the global gauge specified by the
states 
$|\phi_{g}\rangle=e^{iq(k_x+k_y)}(1,1,0)^T$. 
The black and white circles denote $n_{12}=-1$
and~$1$, respectively, whereas a blank implies~$n_{12}=0$. 
It is clear that any of them gives the correct Chern number~$\tilde c_n=-2$.
The field~$n_{12}(k_\ell)$ is gauge-dependent, but their sum is gauge-invariant.
It also shows a convergence of the gauge dependent field $n_{12}$ in the
limit $N_\rmB\to \infty$. 
Although we have fixed the gauge to calculate the field $n_{12}$,
calculations of the Chern  numbers are
performed in {\it any gauge}. We do {\it not\/} need  specific gauge-fixing to
make the gauge connection smooth. An {\it arbitrary\/} gauge (e.g., a phase
choice of eigenvectors given by a numerical library) can be also  
adopted to compute the Chern number.

\begin{figure}[htb]
\begin{center}
\includegraphics[width=0.9\linewidth]{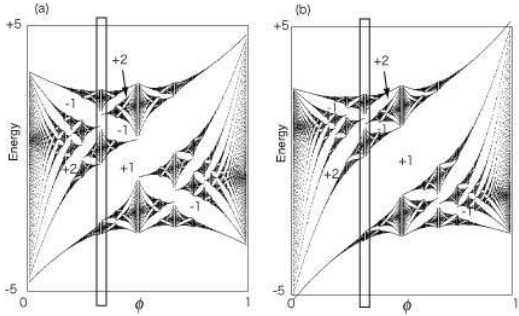}
\end{center}
\caption{
Energy spectra of the NNN models as a function of the flux per
plaquette $\phi$.
(a) $t'=0.18t$ and 
(b) $t'=0.32t$.
Rearrangement of the gap connectivity  is 
observed at the second gap near $\phi=1/3$.
}
\label{f:Nspec}
\end{figure}

\begin{figure}[htb]
\begin{center}
\includegraphics[width=0.99\linewidth]{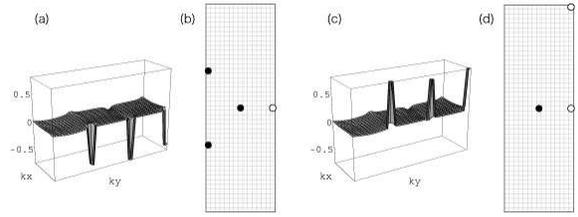}
\end{center}
\caption{
(a):lattice field strength $\tilde F_{12}$ and (b):
configuration of integer field~$n_{12}(k_\ell)$ of the NNN model with
$t'=0.267t$, $N_\rmB=15$.
(c) and (d) the same for parameters $t'=0.268t$.
}
\label{f:Nf1}
\end{figure}

Now let us change a ratio ${t'}/t$ in the NNN model. 
In Figs. \ref{f:Nspec}, we show the energy spectra  of the NNN models.
They are  modified Hofstadter's butterfly diagrams.
The integers in the figures are the Hall conductances when the fermi energy
lies in the gap, that is, it is a sum of the Chern numbers below the energy gap.
Due to the sum rule of the Chern number \cite{Hat045},
the Chern number of the specific energy band is given by 
a difference of two integers,  above and below the band.
As is known~\cite{Hat90},  
the NNN model shows a topological quantum phase transition 
accompanying a discrete change of the Chern numbers.
For example, an increase of ${t'}/t$ causes 
a rearrangement of the  gap connectivity 
near ${t'}/t=0.26$ at $\phi=1/3$.
One can see that the energy gap just above the second band 
at $\phi=1/3$ is rearranged by the small change of ${t'}/t$.
It is associated with a change of the gap labeling from $-1$ to +2.
Since the large energy gap just below the middle band is stable and
the gap is labeled as +1, 
the Chern number of the middle band changes from $-2$ to +1.
This topological change of the Chern number
 is confirmed by a calculation of the 
 gauge dependent field $n_{12}$ in
Figs.\ref{f:Nf1}(b) and (d).
The field strength $\tilde F_{12}$
 just before and the after 
the quantum transition in  also shown in (a) and (c). 
A small change of the parameter ${t'}/t$ does not change 
the spectrum so much. 
On the other hand, 
the field strength $\tilde F_{12}$ is singular
at the critical point.
It is  consistent with the
breakdown of the admissibility as expected. 
That is, the quantum phase transition ( of the present lattice model )
occurs at $ {t'}=t_c$ ($0.267< t_c'/t <0.268$).
This quantum phase transition is associated with a topological 
change of the integer value field $n_{12}$ as shown.

This work was supported in part by Grant-in-Aid for Scientific Research from
JSPS.


\vfill
\begin{thebibliography}{99}

\bibitem{Wen89}
X. G. Wen:
Phys. Rev. {\bf B40}, 7387 (1989).

\bibitem{Hat045}
Y. Hatsugai: 
J. Phys. Soc. Jpn. {\bf 73}, 2604 (2004),
ibid. {\bf 74} 1374 (2005). 


\bibitem{TKNN}
D. J. Thouless, M. Kohmoto, P. Nightingale and M. den Nijs:
Phys. Rev. Lett. {\bf 49}, 405 (1982).


\bibitem{Ber84}
M. V. Berry:
Proc. R. Soc. Lond. {\bf A392}, 45 (1984).

\bibitem{Sim83}
B. Simons: 
Phys. Rev. Lett. {\bf 51}, 2167 (1983).

\bibitem{smf}
T. Senthil, J. B. Marston and M. P. A. Fisher:
Phys. Rev. {\bf B60}, 4245 (1999).

\bibitem{mh}
Y. Morita and Y. Hatsugai:
Phys. Rev. {\bf B62}, 99 (2000).

\bibitem{spin1} 
S. Murakami, N. Nagaosa and S.-C. Zhang:
Science {\bf 5}, 1348 (2003),

\bibitem{spin2} 
J. Sinova, D. Culcer, Q. Niu, N. A. Sinitsyn, T. Jungwirth,
and A.H. MacDonald: 
Phys. Rev. Lett. 92, 126603 (2004).

\bibitem{Hal04}
F. D. M. Haldane:
Phys. Rev. Lett. {\bf 93}, 206602 (2004).

\bibitem{Luscher}
M. L\"uscher:
Commun. Math. Phys. {\bf 85}, 39 (1982) 
Nucl. Phys. {\bf B538} 515 (1999) ,
ibid. {\bf B549}  295 (1999).

\bibitem{Phillips}
A. Phillips:
Ann. Phys. {\bf 161} (1985) 399,
A. Phillips and D. Stone:
Commun. Math. Phys. {\bf 103} (1986) 599;
Commun. Math. Phys. {\bf 131} (1990) 255.





\bibitem{Fujiwara:2000wn}
T. Fujiwara, H. Suzuki and K. Wu:
Prog. Theor. Phys. {\bf 105}, 789 (2001) .

\bibitem{fukui05}
T. Fukui, Y. Hatsugai and H. Suzuki:
J. Phys. Soc. Jpn. {\bf 74}, 1674 (2005). 


\bibitem{wz}
F. Wilczek and A. Zee:
Phys. Rev. Lett. {\bf 52}, 2111 (1984).


\bibitem{Hat90}
Y. Hatsugai and M. Kohmoto:
Phys. Rev. {\bf B42},  8282 (1990).

\end{thebibliography}
\end{document}